\documentstyle[twoside,fleqn,espcrc2]{article}

\def\la{\mathrel{\mathpalette\fun <}}

\def\fun#1#2{\lower3.6pt\vbox{\baselineskip0pt\lineskip.9pt
  \ialign{$\mathsurround=0pt#1\hfil##\hfil$\crcr#2\crcr\sim\crcr}}}
\newcommand{\dd}{\mbox{d}}
\newcommand{\vecc}[1]{\mbox{\boldmath $#1$}}

\newcommand{\AmS}{{\protect\the\textfont2
  A\kern-.1667em\lower.5ex\hbox{M}\kern-.125emS}}

\title{\Large \bf Small--Angle Bhabha Scattering\thanks{
Invited talk presented by L. Trentadue at the Zeuthen
workshop on Elementary Particle Theory ``QCD and QED in Higher
Orders'', Rheinsberg, April 21--26, 1996.}}

\author{A.B. Arbuzov\address{Joint Institute for Nuclear Research
          Dubna, Moscow region, 141980, Russia},
        V.S. Fadin\address{ Budker Institute for Nuclear Physics 
          Novosibirsk State University, 630090, Novosibirsk, Russia},
        E.A.~Kuraev\address{Joint Institute for Nuclear Research
          Dubna, Moscow region, 141980, Russia},
        L.N.~Lipatov\address{St.-Petersburg Institute of Nuclear Physics,
          Gatchina, Leningrad region, 188350, Russia}
        N.P. Merenkov\address{Physics-Techn. Institute, Kharkov, Ukraine}
        and
        L. Trentadue\address{Dipartimento
          di Fisica, Universit\'a di Parma and 
         INFN, Gruppo Collegato di Parma, Parma, Italy.}}

\begin{document}

\begin{abstract}
We present the calculation of
the elastic and inelastic high--energy small--angle electron--positron
scattering with a {\it per mille} accuracy.
\end{abstract}

\maketitle

\section{Introduction}
A considerable attention has been recently
devoted to the Bhabha process \cite{r2,r3,r4}. The
reached accuracy is however still inadequate \cite{r1a}
with respect to the experimental one \cite{r1}. According to these
evaluations the theoretical estimates are still incomplete; moreover,
they are somewhat larger ($\sim$ a factor 2) than
the projected theoretical and experimental precision \cite{r1a} and
are comparable to the currently published experimental precision.

The process that  will be considered  in this work is that of Bhabha scattering
when electrons and positrons are emitted at small angles
with respect to the initial electron and positron directions. We have examined
the
radiative processes inclusively accompanying the main
e$^+$e$^-$$\rightarrow$ e$^+$e$^-$ reaction at
high energies, when both the scattered electron and positron are tagged
within the counter aperture.

We assume that the center--of--mass energies are within the range of the LEP
collider $2\epsilon=\sqrt{s}=90$ --- $200\;$GeV
and the scattering angles are within the range $\theta \simeq 10$ ---
$150\;$mrad.
We assume that the charged--particle detectors have the following polar angle
cuts:

\begin{eqnarray}
&& \theta_1 < \theta_-=\widehat{\vecc{p}_1 \vecc{q}}_1 \equiv
\theta < \theta_3, \nonumber \\
&& \theta_2 < \theta_+=\widehat{\vecc{p}_2 \vecc{q}}_2 < \theta_4,
\nonumber \\ \label{eq2}
&& 0.01\la \theta_i \la 0.1 \;\mbox{rad}\;, 
\end{eqnarray}
where $\vecc{p}_1 , \ \vecc{q}_1\,\; (\vecc{p}_2 , \ \vecc{q}_2\,)$ are
the momenta of the initial and of the scattered electron (positron) in the
center--of--mass frame.

In this talk we present the results of our calculations
of the electron--positron scattering cross--section with an accuracy of
${\cal O }(0.1\%)$.
The squared matrix elements of the various exclusive processes
inclusively contributing to the $e^+e^-\rightarrow e^+e^-$
reaction are integrated in order to define an experimentally measurable
cross--section according to suitable restrictions on the angles and energies
of the detected particles. The different contributions to the electron and
positron distributions, needed for the required accuracy, are presented
using analytical expressions.

In order to define the angular range of interest and the implications on
the required accuracy, let us first briefly discuss, in a general way, the
angle-dependent
corrections to the cross--section.

We consider  $e^+e^-$  scattering at angles as defined in Eq.~(\ref{eq2}).
Within this region, if one expresses the cross--section by means of a series
expansion in terms of angles, the main contribution to the
cross--section
${\dd\sigma}/{\dd\theta^2}$
comes from the diagrams for the scattering amplitudes containing one  exchanged
photon
in the $t$-channel. These diagrams, as well known, show a
singularity of the type $\theta^{-4}$ for $\theta \rightarrow 0$, e.g.
\begin{eqnarray}
\frac{\dd\sigma}{\dd\theta^2} & \sim & \theta^{-4}.\nonumber
\end{eqnarray}

Let us now estimate the correction of order $\theta^2$ to this contribution.
If
\begin{eqnarray}\label{eq3}
 \frac{\dd\sigma}{\dd\theta^2} & \sim & \theta^{-4}(1+c_1\theta^2),
\end{eqnarray}
then, after integration over  $\theta^2$  in the angular range of
Eq.~(\ref{eq2}),
we obtain:

\begin{eqnarray}\label{eq4}
\int \limits_{\theta^2_{\mbox{\tiny min}}}^{\theta^2_{\mbox{\tiny max}}}
\frac{\dd\sigma}{\dd\theta^2}\;\dd\theta^2
& \sim & \theta^{-2}_{\mbox{\tiny min}}(1+c_1\theta^2_{\mbox{\tiny min}}
\ln \frac{\theta^2_{\mbox{\tiny max}}}{\theta^2_{\mbox{\tiny min}}}).
\end{eqnarray}
 We see that, for  $\theta_{\mbox{\tiny min}} = 50\,$mrad and
$\theta_{\mbox{\tiny max}} = 150\;$mrad (we have taken
the case where the $\theta^2$  corrections are maximal), the relative
contribution of the $\theta^2$ terms is about  $5 \times 10^{-3} c_1.$
Therefore,
the terms of relative order $\theta^2$  must be kept only in the
Born cross--section where the coefficient $c_1$ is not small.

A detailed derivation of these results has  been reported elsewhere
(see \cite{r33} and references therein).

\section{Born cross--section and
one-loop virtual and soft corrections}

The Born cross--section for Bhabha scattering within the Standard Model is
well known \cite{r4}. In the small--angle limit is

\begin{eqnarray} \label{eq6}
\frac{\dd\sigma^B}{\theta\dd\theta}=\frac{8 \pi \alpha^2}{\varepsilon^2 \theta^4}
(1-\frac{\theta^2}{2}+\frac{9}{40} \theta^4+\delta_{\mbox{\tiny weak}}),
\end{eqnarray}
where $\varepsilon=\sqrt{s}/2$ is the electron or positron initial energy
and the weak correction term $\delta_{\mbox{\tiny weak}}$, connected with
diagrams with $Z^0$-boson
exchange.

In the pure QED case one-loop radiative corrections to  Bhabha cross--section
were
calculated a long time ago \cite{r5}.
Taking into account a contribution from soft-photon emission with energy less
than a  given finite threshold $\Delta \varepsilon$, we have here for
the cross--section $d \sigma^{(1)}_{QED}$, in the one-loop approximation
for the limiting case of small scattering angles:
\begin{eqnarray}
&& \frac{\dd\sigma^{(1)}_{QED}}{\dd c}=\frac {\dd\sigma^{B}_{QED}}{\dd c}\;
(1-\Pi(t))^{-2}\;(1+\delta),
\\ \nonumber &&
\delta=2 \frac{\alpha}{\pi}\;\biggl[2(1-L)\ln\frac{1}{\Delta}+\frac{3}{2}L
- 2\biggr]  \\ \nonumber &&
+ \frac{\alpha}{\pi}\;\theta^2\;\Delta_{\theta}
+\frac{\alpha}{\pi}\;\theta^2\; \ln\Delta,
\\ \nonumber &&
\Delta_{\theta}=\frac{3}{16}l^2+\frac{7}{12}l-\frac{19}{18}
+\frac{1}{4}\;(\delta_t-\delta_s),
\\ \nonumber &&
\Delta=\frac{\Delta\varepsilon}{\varepsilon},  \qquad l=\ln\frac{Q^2}{s}\simeq
\ln \;\frac {\theta^2}{4}.
\end{eqnarray}
The photon vacuum polarization function $\Pi(t)$  ($\Pi(s)$)  
is defined as follows:
\begin{eqnarray}
\Pi(t)=\frac{\alpha}{\pi} \;\biggl(\delta_t+\frac{1}{3}L-\frac{5}{9}\biggr)
+\frac{1}{4}\;(\frac{\alpha}{\pi})^2 L,
\end{eqnarray}
where
\begin{eqnarray}
L=\ln\frac{Q^2}{m^2},\quad  Q^2=-t=2\varepsilon^2 (1-c),
\end{eqnarray}
and we took into account the leading part of the two--loop contribution
in the
polarization operator. In the Standard Model, $\delta_t$ contains
contributions of muons, tau-leptons, $W$-bosons and hadrons.
For numerical calculations we use for $\Pi(t)$ the results
of Ref.~\cite{r6}.

Taking into account that the large contribution proportional to $\ln\Delta$
disappears when we add the cross--section for the hard emission, one can
verify that terms of relative order $\theta^2$ can be
neglected. Therefore we will omit in higher orders the annihilation diagrams and
multiple-photon exchange diagrams in the scattering channel.
The second simplification is justified by the generalized eikonal
representation for small--angle scattering amplitudes.
In particular, for the case of elastic processes we have  \cite{r7}:

\begin{eqnarray} \label{eq16}
A(s,t)=A_0(s,t)\;F_1^2(t)\;(1-\Pi(t))^{-1} \;\mbox{e}^{i\varphi(t)}\;
\nonumber \\ 
\times \left[1+{\cal {O}}\biggl(\frac{\alpha}{\pi}\;\frac{Q^2}{s}\biggr)
\right],\quad s\gg Q^2\gg m^2,
\end{eqnarray}
where $A_0(s,t)$ is the Born amplitude, $F_1(t)$ is the Dirac form
factor and $\varphi(t)=-\alpha\; \ln(Q^2/\lambda^2)$ is the Coulomb phase,
$\lambda$ is the {\em photon mass\/} auxiliary parameter.
We may consider the eikonal representation as correct
within the required  accuracy\footnote{
Result obtained in paper \cite{r8}, we believe, is incorrect.
It contradicts to the well established result of D.~Yennie et al. \cite{YFS}
about cancelation of infrared singularities.}.

Let us now introduce the dimensionless quantity
$\Sigma=Q_1^2\;\sigma_{\mbox{\tiny exp}}/(4\pi\alpha^2)$,
with $Q_1^2=\varepsilon^2
\theta_1^2$, where  $\sigma_{\mbox{\tiny exp}}$
is the Bhabha--process cross--section
integrated over the typical experimental energy and angular
ranges\footnote{Really this quantity corresponds to some
{\em ideal\/} detectors. It is intended for comparisons
with the results of Monte Carlo event generators.}:
\begin{eqnarray}
\Sigma&=&\frac{Q_1^2}{4\pi\alpha^2}\int\!\dd x_1 \int\!\dd x_2
\;\Theta(x_1x_2-x_c)\nonumber \\
&\times& \int\!\dd^2\vecc{q}_1^{\bot}\;\Theta_1^c\int\!\dd^2\vecc{q}^{\bot}_2
\;\Theta_2^c\;\nonumber \\
&\times& \frac{\dd\sigma^{e^+e^-\rightarrow e^+(\vecc{q}^{\bot}_2,x_2)\,
e^-(\vecc{q}^{\bot}_1,x_1)+X}}{\dd x_1\dd^2\vecc{q}^{\bot}_1\dd x_2
\dd^2\vecc{q}^{\bot}_2}\, ,
\end{eqnarray}
where $x_{1,2}$, $\vecc{q}^{\bot}_{1,2}$  are the energy fractions and the
transverse components of the momenta of the electron and
positron in the final state, $sx_c$ is the experimental cut--off on their
invariant mass squared and the functions
$\Theta_i^c$ do take into account the angular cuts (\ref{eq2}):
\begin{eqnarray}
\Theta_1^c&=&\Theta(\theta_3-\frac{|\vecc{q}^{\bot}_1|}{x_1\varepsilon})\;\;
\Theta(\frac{|\vecc{q}^{\bot}_1|}{x_1\varepsilon}-\theta_1), \nonumber \\
\Theta_2^c&=&\Theta(\theta_4-\frac{|\vecc{q}^{\bot}_2|}{x_2\varepsilon})\;\;
\Theta(\frac{|\vecc{q}^{\bot}_2|}{x_2\varepsilon}-\theta_2).
\end{eqnarray}
 In the case of a symmetrical angular acceptance (we restrict ourselves further
to this case only) we have:
\begin{eqnarray}
\theta_2=\theta_1,\quad \theta_4=\theta_3,\quad
\rho=\frac{\theta_3}{\theta_1} > 1.
\end{eqnarray}

We will present $\Sigma $ as the sum of various contributions:
\begin{eqnarray} \label{eq20}
\Sigma\!\!\!&=&\!\!\!\Sigma_0+\Sigma^{\gamma}+\Sigma^{2\gamma}
+\Sigma^{e^+e^-}+\Sigma^{3\gamma}+\Sigma^{e^+e^-\gamma}\nonumber \\ \nonumber
\!\!\!&=&\!\!\!\Sigma_{00}(1+\delta_{0}+\delta^{\gamma}+\delta^{2\gamma}+
\delta^{e^+e^-}\nonumber \\
\!\!\!&+&\!\!\!\delta^{3\gamma}+\delta^{e^+e^-\gamma}), \quad
\Sigma_{00}=1-\rho^{-2},
\end{eqnarray}
where $ \Sigma_0$ stands for a modified Born contribution, $\Sigma^{\gamma}$
for a contribution of one-photon emission (real and virtual) and so on.
The values of the $\delta^i$ as function of $x_c$ are given in Table~1
(see below).  We shall
slightly modify the perturbation theory, using the full propagator for the
$t$-channel
photon, which takes into account the growth of the electric charge at small
distances.
By integrating with this convention, we obtain:
\begin{equation}
\Sigma_0=\theta_1^2\int\limits_{\theta_1^2}^{\theta_2^2}
\frac{\dd\theta^2}{\theta^4}
(1-\Pi(t))^{-2}+\Sigma_W+\Sigma_\theta,
\end{equation}
where $\Sigma_W$ is the correction due to the weak interaction:
\begin{equation}
\Sigma_W=\theta_1^2\int\limits_{\theta_1^2}^{\theta_2^2}\frac{\dd\theta^2}
{\theta^4}\delta_{\mbox{\tiny weak}}\, ,
\end{equation}
and the term $\Sigma_{\theta}$ comes from the expansion of the Born
cross--section in powers of $\theta^2$,
\begin{equation}
\Sigma_{\theta}=\theta_1^2\int\limits_{1}^{\rho^2}\!\frac{\dd z}{z
(1-\Pi(-zQ_1^2))^2}\biggl(-\frac{1}{2}+z\theta_1^2\frac{9}{40}\biggr).
\end{equation}
The remaining contributions to $\Sigma$ in (\ref{eq20}) are considered below.

\subsection{Single hard-photon emission }

In order to calculate the contribution to $\Sigma$ due to the hard-photon
emission we start from  the corresponding differential cross--section
written in terms of energy fractions $x_{1,2}$ and transverse components
$\vecc{q}^{\bot}_{1,2}$ of the final particle momenta \cite{r9}:
\begin{eqnarray} \label{eq24}
&& \frac{\dd\sigma_{B}^{e^+e^-\rightarrow e^+e^-\gamma}}
{\dd x_1\dd^2\vecc{q}^{\bot}_1\dd x_2\dd^2\vecc{q}^{\bot}_2}
=\frac{2\alpha^3}{\pi^2} (1+{\cal{O}}(\theta^2)) \nonumber \\
&&\ \times \biggl\{\frac{R(x_1;\vecc{q}^{\bot}_1,
\vecc{q^{\bot}}_2)
\;\delta(1-x_2)}{(\vecc{q}^{\bot}_2)^4\;(1-\Pi(-(\vecc{q}^{\bot}_2)^2))^2}
 \nonumber \\
&&\ + \frac{R(x_2;\vecc{q}^{\bot}_2,\vecc{q}^{\bot}_1)\;
\delta(1-x_1)}{(\vecc{q}^{\bot}_1)^4\;
(1-\Pi(-(\vecc{q}^{\bot}_1)^2))^2}\biggr\},
\end{eqnarray}
where
\begin{eqnarray} \label{eq25}
&& R(x;\vecc{q}^{\bot}_1,\vecc{q}^{\bot}_2)=\frac{1+x^2}{1-x}\;
\biggl[\frac{(\vecc{q}^{\bot}_2)^2(1-x)^2}{d_1d_2} \nonumber \\ 
&&\ -\frac{2m^2(1-x)^2x}{1+x^2}\;\frac{(d_1-d_2)^2}{d_1^2d_2^2}\biggr],
\\ \nonumber 
&& d_1=m^2(1-x)^2+(\vecc{q}^{\bot}_1-\vecc{q}^{\bot}_2)^2,
\\ \nonumber  
&& d_2=m^2(1-x)^2+(\vecc{q}^{\bot}_1-x\vecc{q}^{\bot}_2)^2,
\end{eqnarray}
and we use the full photon propagator for the $t$-channel
photon.
Performing a simple azimuthal angle integration of Eq.~(\ref{eq24}) 
we obtain for the hard-photon emission the contribution $\Sigma^H$:
\begin{eqnarray} \label{eq30}
&&\!\!\!\Sigma^H=\frac{\alpha}{\pi} \int\limits_{x_c}^{1-\Delta}\dd x
\frac{1+x^2}{1-x} \int\limits_{1}^{\rho^2}
\frac{\dd z}{z^2(1-\Pi(-zQ_1^2))^2} \nonumber \\
&&\!\!\!\times \biggl\{ [1 +\Theta(x^2\rho^2-z)]\;
(L-1)+k(x,z) \biggr\},\nonumber
\\ 
&&\!\!\! k(x,z)= \frac{(1-x)^2}{1+x^2} \;[1+\Theta(x^2\rho^2-z)]+ \; L_1
\nonumber \\
&&\!\!\! +\Theta(x^2\rho^2-z)\;L_2
+\Theta(z-x^2\rho^2) L_3 \, ,
\end{eqnarray}
where $L=\ln(zQ_1^2/m^2)$ and
\begin{eqnarray} \label{l123}
L_1&=&\ln\left|\frac{x^2(z-1)(\rho^2-z)}{(x-z)(x\rho^2-z)}\right|, \nonumber \\
L_2&=&\ln\left|\frac{(z-x^2)(x^2\rho^2-z)}{x^2(x-z)(x\rho^2-z)}\right|,
\nonumber \\ 
L_3&=&\ln\left|\frac{(z-x^2)(x\rho^2-z)}{(x-z)(x^2\rho^2-z)}\right|.
\end{eqnarray}
It is seen from Eq.~(\ref{eq30}) that $\Sigma^H$ contains  the  auxiliary parameter
$\Delta$. This parameter disappears, as it should, in the sum
$\Sigma^{\gamma} = \Sigma^H+\Sigma^{V+S}$, where $\Sigma^{V+S}$ is the
contribution of virtual and soft real photons:

\begin{eqnarray} \label{eq32}
\Sigma^{\gamma}&=&\frac{\alpha}{\pi}\int\limits_{1}^{\rho^2}
\frac{\dd z}{z^2}\int\limits_{x_c}^{1}
\dd x (1-\Pi(-zQ_1^2))^{-2}\; \nonumber \\
& \times &\biggl\{(L-1) P(x)
[1+\Theta(x^2\rho^2-z)] \nonumber \\
&+&\frac {1+x^2}{1-x}k(x,z)-\delta(1-x) \biggr\},
\end{eqnarray}
where is the non--singlet splitting kernel.

\section{Radiative corrections to ${\cal{O}}(\alpha^2)$}

A systematic treatment of all ${\cal O}(\alpha^2)$ contributions
is absent up to now. This is mainly due to the extreme complexity of
the analysis (more then 100 Feynman diagrams are to be taken into
account considering elastic and inelastic processes).
Nevertheless in the case of small scattering angles we may restrict
ourselves by considering only diagrams of the scattering type.

\subsection{Virtual and soft corrections
to the hard-photon emission}

By evaluating the corrections arising from the emission of virtual and real
soft photons which accompany a single hard-photon we will consider two cases.
The first case corresponds to the emission of the photons by the same fermion.
The second one occurs when the hard-photon is emitted by another fermion:

\begin{eqnarray} \label{eq46}
\dd\sigma\bigg|_{H(S+V)}&=&\dd\sigma^{H(S+V)} + \dd\sigma_{H(S+V)}\nonumber \\
&+& \dd\sigma^{H}_{(S+V)} + \dd\sigma_{H}^{(S+V)}.
\end{eqnarray}
In the case when both fermions emit, one finds that:
\begin{eqnarray} \label{eq47}
\Sigma^{H}_{(S+V)}+\Sigma_{H}^{(S+V)}&=&2\Sigma^{H}
\bigl(\frac{\alpha}{\pi}\bigr)\biggl[(L-1)\ln\Delta \nonumber \\
&+& \frac{3}{4}L-1\biggr],
\end{eqnarray}
where $\Sigma^H$ is given in Eq.~(\ref{eq30}).
A more complex expression arises when the radiative corrections
are applied to the same fermion line.
Here the cross--section may be expressed in terms
of the Compton tensor with an off--shell photon~\cite{r11}, which
describes the process
\begin{eqnarray} \label{eq48}
\gamma^{*}(q) + e^-(p_1)\rightarrow e^-(q_1) + \gamma(k)
+ (\gamma_{\mbox{\tiny soft}}).
\end{eqnarray}

The result has the form:
\begin{eqnarray} \label{eq49}
&&\!\!\!\!\! \Sigma^{H(S+V)} = \Sigma_{H(S+V)}= \frac{1}{2}({\alpha \over \pi})^2
\int\limits_1^{\rho^2}\frac{\dd z}{z^2} \nonumber \\
&&\!\!\!\!\! \times \int\limits_{x_c}^{1-\Delta}\frac{\dd x(1+x^2)}{1-x}\;L
\;\biggl\{\biggl(2\ln\Delta - \ln x
+ \frac{3}{2}\biggr)  \nonumber \\
&&\!\!\!\!\! \times [(L-1)(1+\Theta)+k(x,z)]
+\frac{1}{2}\ln^2x  \nonumber \\
&&\!\!\!\!\! + (1+\Theta)[-2+\ln x-2\ln\Delta]+(1-\Theta) \nonumber \\ 
&&\!\!\!\!\! \times \biggl[\frac{1}{2} L \ln x
+ 2\ln\Delta \ln x-\ln x\ln(1-x)-\ln^2x \nonumber \\ 
&&\!\!\!\!\! - \mbox{Li}_2(1-x) - \frac{x(1-x)+4x\ln x}{2(1+x^2)}\biggr] 
\\ \nonumber
&&\!\!\!\!\! - \frac{(1-x)^2}{2(1+x^2)}\biggr\}, 
\quad \mbox{Li}_2(x)\equiv -\int\limits_{0}^{x}\frac{\dd t}{t}\ln(1-t),
\end{eqnarray}
where $k(x,z)$ is given in Eq.~(\ref{eq30}) and
$\Theta \equiv \Theta (x^2\rho^2-z)$.

\subsection{Double hard-photon bremsstrahlung}

We now consider the contribution given by the process of emission of two hard
photons. We will distinguish two cases:  a) the double simultaneous
bremsstrahlung in opposite directions along electron and positron momenta,
and b) the double bremsstrahlung in the same direction along electron or
positron momentum.
The differential cross--section in the first case can be obtained by
using the factorization property of cross--sections within the impact parameter
representation \cite{r12}. It takes the following form \cite{r9}:
\begin{eqnarray} \label{eq50}
&&\!\!\!\! \frac{\dd\sigma^{e^+e^-\rightarrow (e^+\gamma)(e^-\gamma)}}
{\dd x_1\dd^2\vecc{q}^{\bot}_1\dd x_2\dd^2\vecc{q}^{\bot}_2}
=\frac{\alpha^4}{\pi^3}\int\limits_{}^{} 
\frac{\dd^2\vecc{k}^{\bot}}{\pi (\vecc{k}^{\bot})^4}\;\nonumber \\
&&\!\!\!\! \times \frac{R(x_1;\vecc{q}^{\bot}_1,\vecc{k}^{\bot})
R(x_2;\vecc{q}^{\bot}_2,-\vecc{k}^{\bot})}{(1-\Pi(-(\vecc{k}^{\bot})^2))^{2}},
\end{eqnarray}
where $R(x;\vecc{q}^{\bot},\vecc{k}^{\bot})$ is given by Eq.~(\ref{eq25}).
The calculation of the corresponding contribution $ \Sigma^H_H $ to $ \Sigma $
is analogous to the case of the single
hard-photon emission and the result has the form:
\begin{eqnarray} \label{eq52}
&&\!\!\!\!\! \Sigma_H^H=\frac{1}{4}\bigl(\frac{\alpha}{\pi}\bigr)^2\!\!
\int\limits_{0}^{\infty}\!
\frac{\dd z}{z^2(1-\Pi(-zQ_1^2))^2}\!\int\limits_{x_c}^{1-\Delta}\!\!\!\dd x_1
\nonumber \\
&&\!\!\!\!\! \times\!\! \int\limits_{x_c/x_1}^{1-\Delta}\!\!\!\dd x_2\;
\frac{1+x_1^2}{1-x_1}\;\frac{1+x_2^2}{1-x_2}\;\Phi(x_1,z)\Phi(x_2,z),
\end{eqnarray}
where (see Eq.~(\ref{l123})):
\begin{eqnarray} \label{eq53}
&&\!\!\!\! \Phi(x,z)=(L-1)[\Theta(z-1)\Theta(\rho^2-z)
\nonumber \\ \nonumber
&&\!\!\!\! +\Theta(z-x^2)\Theta(\rho^2x^2-z)]
\\ \nonumber
&&\!\!\!\! +L_3[-\Theta(x^2-z)+\Theta(z-x^2\rho^2)]
\\ \nonumber 
&&\!\!\!\! +\biggl(L_2+\frac{(1-x)^2}{1+x^2}\biggr)
\Theta(z-x^2)\Theta(x^2\rho^2-z) 
\\ \nonumber
&&\!\!\!\! + \biggl(L_1+\frac{(1-x)^2}{1+x^2}\biggr)
\Theta(z-1)\Theta(\rho^2-z) \\ \nonumber
&&\!\!\!\!\! + (\Theta(1-z)\! -\!\Theta(z-\rho^2))
\ln\left|\frac{(z-x)(\rho^2-z)}{(x\rho^2-z)(z-1)}\right|.
\end{eqnarray}

Let us now turn to  the double hard-photon emission in the
same direction and the hard e$^+$ e$^-$ pair production.
Here we use the method  developed by one of us \cite{r13,r14}.
We will distinguish the
collinear and semi--collinear kinematics of final particles.
In the first case all produced particles move in the cones
within the polar angles $\theta_i<
\theta_0\ll 1$ centered along the charged-particle momenta (final or initial).
In the semi--collinear region only one produced particle moves inside
those cones,
while the other moves outside them. For the totally inclusive
cross--section, such a distinction no longer has physical meaning and the
dependence on the auxiliary parameter $\theta_0$ disappears.
We underline that in this way all double and single--logarithmical
contributions may be extracted rigorously. The contribution of the
region when both the photons move outside the small cones does not
contain any large logarithm $L$. The systematic omission of those
contributions in the double bremsstrahlung and pair production processes
is the source of uncertainties of order
$(\alpha/\pi)^2\leq 0.6\cdot 10^{-5}$.

The contribution of both collinear and semi--collinear regions
(we consider for definiteness
the emission of both hard photons along the electron, since the
contribution of the emission along the positron is the same) has
the form:
\begin{eqnarray} \label{eq54} \nonumber
&&\!\!\!\!\! \Sigma^{HH}=\Sigma_{HH}=\frac{1}{4}\bigl(\frac{\alpha}{\pi}\bigr)^2
\int\limits_{1}^{\rho^2}\frac{\dd z}{z^2(1-\Pi(-zQ_1^2))^2} \\ \nonumber
&&\!\!\!\!\! \times \int\limits_{x_c}^{1-2\Delta}\!\!\dd x
\int\limits_{\Delta}^{1-x-\Delta}\!\!\!\dd x_1
\frac{I^{HH}L}{x_1(1-x-x_1)(1-x_1)^2}, \\ \nonumber
&&\!\!\!\!\! I^{HH}=A\;\Theta(x^2\rho^2-z)+B
\\ 
&&\!\!\!\!\! +C\;\Theta((1-x_1)^2\rho^2-z),
\end{eqnarray}
where $A$, $B$ and $C$ are known functions \cite{r33}.

The total expression $\Sigma^{2\gamma}$, which describes the
contribution to (\ref{eq20}) from the two--photon (real and virtual)
 emission processes reads as follows:
\begin{eqnarray} \label{eq57}
\Sigma^{2\gamma}&\!\! =\!\!&\Sigma_{S+V}^{\gamma\gamma}+2\Sigma^{H(V+S)}
+ 2\Sigma^{H}_{S+V} 
\nonumber \\
&\!\! +\!\!& \Sigma_H^H+2\Sigma^{HH} \\ \nonumber
&\!\! =\!\!& \Sigma^{\gamma\gamma}+\;\Sigma_{\gamma}^{\gamma}+
(\frac{\alpha}{\pi})^2{\cal{L}}(\phi^{\gamma\gamma}
+\phi^{\gamma}_{\gamma}), \ \ 
\\ \nonumber 
{\cal{L}}&\!\! = \!\!& \ln\frac{\varepsilon^2\theta_1^2}{m^2}.
\end{eqnarray}
The leading contributions $ \Sigma^{\gamma\gamma},
\Sigma_{\gamma}^{\gamma} $ have the following forms:
\begin{eqnarray} \label{eq58}
\!\! &&\!\!\!\!\!\! \Sigma^{\gamma\gamma}=\frac{1}{2}
\bigl(\frac{\alpha}{\pi}\bigr)^2
\int\limits_{1}^{\rho^2}\frac{\dd z}{z^2}L^2
(1-\Pi(-Q_1^2z))^{-2}
\nonumber \\ \nonumber
\!\! &&\!\!\!\!\!\! \times \int\limits_{x_c}^{1}\dd x\;\biggl\{
\frac{1}{2}P^{(2)}(x)\;[\;\Theta(x^2\rho^2-z)+1] \\ 
\!\! &&\!\!\!\!\!\! + \int\limits_{x}^{1}\frac{\dd t}{t}
P(t)\;P(\frac{x}{t})\;\Theta(t^2\rho^2-z)\biggr\}, \\  \label{eq59}
\!\! &&\!\!\!\!\!\! \Sigma_{\gamma}^{\gamma}
=\frac{1}{4}\bigl(\frac{\alpha}{\pi}\bigr)^2
\int\limits_{0}^{\infty}\frac{\dd z}{z^2}L^2
(1-\Pi(-Q_1^2z))^{-2}
\nonumber \\ \nonumber
\!\! &&\!\!\!\!\!\! \times \int\limits_{x_c}^{1}\dd x_1
\int\limits_{x_c/x_1}^{1}\dd x_2 P(x_1)P(x_2) \\ \nonumber
\!\! &&\!\!\!\!\!\! \times \bigl[\Theta(z-1)\Theta(\rho^2-z)
+ \Theta(z-x_1^2)\Theta(x_1^2\rho^2-z)\bigr] \\ \nonumber
\!\! &&\!\!\!\!\!\! \times \bigl[\Theta(z-1)\Theta(\rho^2-z)
+ \Theta(z-x_2^2)\Theta(x_2^2\rho^2-z)\bigr].
\end{eqnarray}

We see that the leading contributions to $\Sigma^{2\gamma}$ may be expressed
in terms of kernels for the evolution equation for structure functions.

The functions $\phi^{\gamma\gamma}$ and $\phi^{\gamma}_{\gamma}$ 
in expression Eq.~(\ref{eq57}) collect the next-to-leading
contributions which cannot be obtained by the
structure functions method \cite{r15}. They have a form that can be obtained
by comparing the results in the leading logarithmic approximation
with the logarithmic ones given above.

\section{Pair production}

Pair production process in high--energy e$^+$ e$^-$ collisions
was considered
about 60 years ago (see \cite{r9} and references therein).
In particular it was found that the total cross--section
contains cubic terms in large logarithm $L$. These terms come
from the kinematics when the scattered electron and positron
move in narrow (with opening angles $\sim m/\epsilon$) cones
and the created pair have the invariant mass of the order
of $m$ and moves preferably along either the electron
beam direction or the positron one. According to the conditions
of the LEP detectors, such a kinematics can be excluded.
In the relevant kinematical region a parton-like description
could be used giving $L^2$ and $L$-enhanced terms.

We accept the LEP~1 conventions whereby an event of the Bhabha process
is defined as one in which
the angles of the simultaneously registered particles hitting
opposite detectors (see Eq.~(\ref{deli0})).

The method, developed by one of us (N.P.M.) \cite{r13,r14},
of calculating the real hard pair production cross--section
within logarithmic accuracy (see the discussion in sect.~6)
consists in separating the
contributions of the collinear and semi--collinear kinematical regions.
In the first one (CK)  we suggest that both electron and positron from
the created pair go in the narrow cone around the direction of one
charged particle [the projectile (scattered) electron $\vecc{p}_1$
$(\vecc{q}_1)$ or the projectile (scattered) positron $\vecc{p}_2$
$(\vecc{q}_2)]$:
\begin{eqnarray} \label{p1}
&& \widehat{\vecc{p}_+\vecc{p}_-} \sim \widehat{\vecc{p}_-\vecc{p}_i}
\sim \widehat{\vecc{p}_+\vecc{p}_i}
< \theta_0 \ll 1, \nonumber \\
&& \varepsilon\theta_0/m \gg 1, 
\vecc{p}_i=\vecc{p}_1,\,\vecc{p}_2,\,\vecc{q}_1,\,\vecc{q}_2\, .
\end{eqnarray}
The contribution of the CK contains terms of order
$(\alpha L/\pi)^2$, $(\alpha/\pi)^2L\ln(\theta_0/\theta)$
and $(\alpha/ \pi)^2L$, where $\theta=\widehat{\vecc{p}_-\vecc{q}_1}$ is
the scattering angle.
In the semi--collinear region
only one of conditions (\ref{p1}) on the angles is fulfilled:
\begin{eqnarray} \label{p2}
&& \widehat{\vecc{p}_+\vecc{p}_-} < \theta_0,\quad
\widehat{\vecc{p}_{\pm}\vecc{p}_i} > \theta_0\, ;
\nonumber \\
\mbox{or} && \quad
\widehat{\vecc{p}_-\vecc{p}_i} < \theta_0,\quad
\widehat{\vecc{p}_+\vecc{p}_i} > \theta_0\, ; \\ \nonumber
\mbox{or} && \quad
\widehat{\vecc{p}_-\vecc{p}_i} > \theta_0,\quad
\widehat{\vecc{p}_+\vecc{p}_i} < \theta_0\, .
\end{eqnarray}
The contribution of the SCK contains terms of the form:
\begin{eqnarray} \label{p3}
\left( \frac{\alpha}{\pi} \right) ^2 L \ln\frac{\theta_0}{\theta},
\qquad \left( \frac{\alpha}{\pi} \right) ^2 L.
\end{eqnarray}
The auxiliary parameter $\theta_0$ drops out in the total sum of
the CK and SCK contributions.

Taking into account the leading and
next-to-leading terms we can write the full hard pair contribution
including also the pair emission along the positron direction,
after the integration over $x_2$ as
\begin{eqnarray} \label{eq:41}
\sigma_{\mbox{\tiny hard}}&\!\!\! =\!\!\!&2\;\frac{\alpha^4}{\pi Q_1^2}\int\limits_{1}^{\rho^2}
\frac{\dd z}{z^2} \int\limits_{x_c}^{1-\Delta}\!\dd x
\biggl\{ L^2(1+\Theta)R(x) 
\nonumber \\
&\!\!\! +\!\!\!& {\cal L}[\Theta F_1(x)+F_2(x)] \biggr\},
\\ \nonumber
F_{1,2}(x)&\!\!\! =\!\!\!&d(x)+C_{1,2}(x),
\\ \nonumber
d(x)&\!\!\! =\!\!\!&\frac{1}{1-x}
\biggl(\frac{8}{3}\ln(1-x)-\frac{20}{9}\biggr),
\\ \nonumber
R(x)&\!\!\! =\!\!\!&\frac{1}{3} \frac{1+x^2}{1-x} + \frac{1-x}{6x}(4+7x+4x^2)
\nonumber \\ \nonumber
&\!\!\! +\!\!\!& (1+x)\ln x,
\end{eqnarray}
where $C_1(x)$ and $C_2(x)$ are known functions \cite{r33}.

Eq.~(\ref{eq:41}) describes the small--angle
high--energy cross--section
for the pair production process, provided that the created hard pair 
can move along both electron and positron beam directions. 

The contribution to the cross--section of the small--angle Bhabha scattering
connected with the real soft (with energy lower than
$\Delta  \varepsilon$) and virtual pair production can be defined
\cite{r18} by the formula:
\begin{eqnarray} \label{eq:45}
\sigma_{\mbox{\tiny soft+virt}}&=&\frac{4\alpha^4}{\pi Q_1^2}
\int\limits_{1}^{\rho^2}\;\;\frac{\dd z}{z^2} \biggl\{L^2
\biggl(\frac{2}{3}\ln\Delta + \frac{1}{2}\biggr)
\nonumber \\
&+& {\cal L}\biggl(-\frac{17}{6}+\frac{4}{3}\ln^2\Delta
\nonumber \\
&-& \frac{20}{9} \ln \Delta - \frac{4}{3} \zeta_2\biggr) \biggr\}.
\end{eqnarray}
Using Eqs.~(\ref{eq:41}) and (\ref{eq:45}) it is easy to verify
that the auxiliary parameter $\Delta$ is cancelled in the sum
$\sigma_{\mbox{\tiny pair}}=\sigma_{\mbox{\tiny hard}}
+\sigma_{\mbox{\tiny soft+virt}}$.
We can, therefore, write the total contribution $\sigma_{\mbox{\tiny pair}}$ as
\begin{eqnarray} \label{eq:46}
&& \sigma_{\mbox{\tiny pair}}=\frac{2\alpha^4}{\pi Q_1^2}
\int\limits_{1}^{\rho^2}\;\;
\frac{\dd z}{z^2} \biggl\{ L^2\bigl(1+\frac{4}{3}\ln(1-x_c)
\nonumber \\
&& -\frac{2}{3}\int\limits_{x_c}^{1}\frac{\dd x}{1-x}\bar{\Theta}\bigr)
+ {\cal L}\biggl[-\frac{17}{3}- \frac{8}{3}\zeta_2
\\ \nonumber
&& -\frac{40}{9}\ln(1-x_c)+\frac{8}{3}\ln^2(1-x_c)
\\ \nonumber 
&&+ \int\limits_{x_c}^{1}\frac{\dd x}{1-x}\bar{\Theta}\cdot
\bigl(\frac{20}{9}-\frac{8}{3}\ln(1-x)\bigr) \biggr]
\\ \nonumber 
&& + \int\limits_{x_c}^{1}\;\!\dd x\bigl[ L^2(1+\Theta)\bar{R}(x)
+ {\cal L}(\Theta C_1(x) 
\\  
&& + C_2(x)) \bigr] \biggr\}, \\ \nonumber
&& \bar{R}(x)=R(x)-\frac{2}{3(1-x)},
\quad
\bar{\Theta}=1-\Theta.
\end{eqnarray}

The right-hand side of Eq.~(\ref{eq:46}) gives the contribution to
the small--angle Bhabha scattering cross--section for
pair production. It is finite and can be used for numerical
estimations. The leading term can be described by the
electron structure function $D_e^{\bar{e}}(x)$ \cite{r16}.

\section{Terms of ${\cal O}(\alpha{\cal{L}})^3 $ }

In order to evaluate the leading logarithmic contribution
represented by terms of the type $(\alpha {\cal {L}})^3$,
we use the iteration up to $\beta^3 $ of the
master equation~\cite{r15} obtained in Ref.~\cite{r16}.
To simplify the analytical
expressions we adopt here a realistic assumption about the smallness
of the threshold for the detection of the hard subprocess energy
and neglect terms of the order of:
\begin{eqnarray} \label{q1}
x_c^n(\frac{\alpha}{\pi}{\cal L})^3 \leq 3\cdot 10^{-5},\qquad
n=1,2,3\, .
\end{eqnarray}
We may, therefore, limit ourselves to consider the emission by the
initial electron and positron. Three photons (virtual and real)
contribution to $\Sigma$ have the form:
\begin{eqnarray} \label{q2}
\!\! &&\!\!\!\!\! \Sigma^{3\gamma}=\frac{1}{4}
(\frac{\alpha}{\pi}{\cal{L}})^3\!
\int\limits_{1}^{\rho^2}\!\frac{\dd z}{z^2}
\int\limits_{x_c}^{1}\!\!\dd x_1 \!
\int\limits_{x_c}^{1}\!\!\dd x_2\;\Theta(x_1x_2-x_c)\;
\nonumber \\
\!\! &&\!\!\!\!\! \times \biggl[\frac{1}{6}\delta(1-x_2)\;P^{(3)}(x_1)\Theta(x_1^2\rho^2-z)
\nonumber \\
\!\! &&\!\!\!\!\! +\frac{1}{2x_1^2}P^{(2)}(x_1)P(x_2)\Theta_1\Theta_2\biggr]\;(1+{\cal O}(x_c^3)),
\end{eqnarray}
\begin{eqnarray} \label{q3}
\Theta_1\Theta_2&=&\Theta\bigl(z-\frac{x_2^2}{x_1^2}\bigr)\;
\Theta\bigl(\rho^2 \frac{x_2^2}{x_1^2}-z\bigr).
\end{eqnarray}
$P^{(3)}(x)$ is the three-loop splitting function.
The contribution to $\Sigma$ of the process of pair production
accompanied by photon
emission when both, pair and photons, may be real and virtual has the
form (with respect to paper by M.~Skrzypek~\cite{r16}
we include also the non--singlet mechanism
of pair production):
\begin{eqnarray}
&& \Sigma^{e^+e^-\gamma}=\frac{1}{4}(\frac{\alpha}{\pi}
{\cal{L}})^3\int\limits_{1}^{\rho^2}\,
\dd z\;z^{-2}\int\limits_{x_c}^{1}\,\dd x_1\int\limits_{x_c}^{1}\,\dd x_2
\nonumber \\
&&\times \Theta(x_1x_2-x_c)\;
\{\frac{1}{3}[R^P(x_1)-\frac{1}{3}R^s(x_1)]
\nonumber \\
&&\times \delta(1-x_2)\Theta(x_1^2\rho^2-z)
\nonumber \\
&&+\frac{1}{2x_1^2}\;P(x_2)R(x_1)
\;\Theta_1\Theta_2\},
\end{eqnarray}
where
\begin{eqnarray}
&& R(x)=R^s(x)+\frac{2}{3}P(x), 
\nonumber \\
&& R^s(x)=\frac{1-x}{3x}(4+7x+4x^2)+2(1+x)\ln x,
\nonumber \\
&& R^P(x)=R^s(x)(\frac{3}{2}+2\ln(1-x))
\nonumber \\
&& +(1+x)(-\ln^2x+ 4\mbox{Li}_2(1-x)  
\nonumber \\
&& +\frac{1}{3}(-9-3x+8x^2)\ln x
\nonumber \\
&& +\frac{2}{3}(-\frac{3}{x}-8+8x+3x^2) + \frac{2}{3}P^{(2)}(x).
\end{eqnarray}
The quantity $ \Sigma $ depends on the parameters $x_c,\rho$
and $Q_1^2$.

\section{Estimates of neglected terms and numerical results}

The uncertainty of our calculations is defined by neglected terms.
Let us list them.

a) Terms of the first order RC coming from annihilation--type
diagrams (15):
\begin{eqnarray}
\frac{\alpha}{\pi}\theta_1^2\int\limits_{\theta_1^2}^{\theta_2^2}
\frac{\dd \theta}{\theta^2}\;\Delta_\theta \leq 0.10\cdot 10^{-4}.
\end{eqnarray}

b) Similar terms in the second order do not exceed
(see sect.~4)
\begin{eqnarray}
&& (\frac{\alpha}{\pi})^2\theta_1^2
\int\limits_{\theta_1^2}^{\theta_2^2}\frac{\dd \theta}{\theta^2}\; l^4
\leq 0.23\cdot 10^{-4},\\ \nonumber
&& (\frac{\alpha}{\pi})^2(\theta_2^4-\theta_1^4){\cal L}^4\leq
0.5\cdot 10^{-5}.
\end{eqnarray}

c) We neglect terms which violate the eikonal approximation:
\begin{eqnarray}
\frac{\alpha}{\pi}\;\frac{Q^2}{s}\leq 0.3\cdot 10^{-6}.
\end{eqnarray}

d) We omit term of the second order which are not enhanced by
large logarithms:
\begin{eqnarray}
(\frac{\alpha}{\pi})^2= 0.5\cdot 10^{-5}.
\end{eqnarray}

e) Creation of heavy pairs $(\mu\mu$, $\tau\tau$, $\pi\pi$, $\dots)$ gives
in sum at least one order of magnitude smaller than the corresponding
contribution due to light particle production~\cite{rele}:
\begin{eqnarray}
\Sigma_{\pi\pi}+\Sigma_{\mu\mu}+\Sigma_{\tau\tau} \leq 0.1\;\Sigma^{e^+e^-}
\leq 0.5\cdot 10^{-4}.
\end{eqnarray}

f) Higher--order corrections, including soft and collinear multi-photon
contributions, can be neglected since they only give contributions
of the type $(\alpha L/\pi)^4\leq 0.2\cdot 10^{-5}$ or less.

g) The terms in the third order associated with the emission off
the final particles\footnote{Usually, in a calorimetric experimental
set--up such terms do not contribute.}:
\begin{eqnarray}
x_c(\frac{\alpha {\cal L}}{\pi})^3 \leq 0.3\cdot 10^{-4}\ \ \
(\mbox{for}\ x_c=0.5).
\end{eqnarray}

Regarding all the uncertainties a)--g) and (82) as independent ones
we conclude the total theoretical uncertainty of our results to be
$\pm 0.006\%$. "

Let us  define $\Sigma_0^0$ to be equal to $\Sigma_0|_{\Pi=0}$
(see Eq.~(21)), which corresponds to the Born cross--section
obtained by switching off the vacuum polarization contribution
$\Pi(t)$. For the experimentally observable cross--section we obtain:

\begin{eqnarray}\label{deli0}
\sigma&=&\frac{4\pi\alpha^2}{Q_1^2}\Sigma_0^0\;(1+\delta_0+\delta^{\gamma}
+\delta^{2\gamma}+\delta^{e^+e^-}
\nonumber \\
&+& \delta^{3\gamma}+\delta^{e^+e^-\gamma}),
\end{eqnarray}
where
\begin{eqnarray}
\Sigma_0^0=\Sigma_0|_{\Pi=0}=1-\rho^{-2}+\Sigma_W+\Sigma_{\theta}|_{\Pi=0}
\end{eqnarray}
and
\begin{eqnarray}
\delta_0=\frac{\Sigma_0-\Sigma_0^0}{\Sigma_0^0},\;
\delta^{\gamma}=\frac{\Sigma^{\gamma}}{\Sigma_0^0},\;
\delta^{2\gamma}=\frac{\Sigma^{2\gamma}}{\Sigma_0^0},\;
\cdots\; .
\end{eqnarray}

The numerical results are presented in Table~1.
\begin{table*}[hbt]
\setlength{\tabcolsep}{.8pc}
\newlength{\digitwidth} \settowidth{\digitwidth}{\rm 0}
\catcode`?=\active \def?{\kern\digitwidth}
\caption{The values of $\delta^i$ in per cent
for  $s^{1/2}=$91.161~GeV, $\theta_1=1.61^{\circ}$,
$\theta_2=2.8^{\circ}$, $\sin^2\theta_W=0.2283$,
$\Gamma_Z=2.4857$ GeV.}
\label{tab:1}
\begin{tabular*}{155mm
}{@{}l@{\extracolsep{
4.pt 
}}|r|r|r|r|r|r|r|r|r|}
\hline
$x_c$ & $\delta_0$ & $\delta^{\gamma} $
&$\delta^{2\gamma}_{\mbox{\tiny leading}}$
&$\delta^{2\gamma}_{\mbox{\tiny nonlead}}$
& $\delta^{e^+e^-} $ & $\delta^{e^+e^-\gamma} $
&$\delta^{3\gamma} $ &$\sum \delta^i $ \\ \hline
0.1& 4.120& $-$8.918& 0.657&  0.162& $-$0.016& $-$0.017
& $-$0.019& $-$4.031 \\
0.2& 4.120& $-$9.226& 0.636&  0.156& $-$0.027& $-$0.011
& $-$0.016& $-$4.368 \\
0.3& 4.120& $-$9.626& 0.615&  0.148& $-$0.033& $-$0.008
& $-$0.013& $-$4.797 \\
0.4& 4.120&$-$10.147& 0.586&  0.139& $-$0.039& $-$0.005
& $-$0.010& $-$5.356 \\
0.5& 4.120&$-$10.850& 0.539&  0.129& $-$0.044& $-$0.003
& $-$0.006& $-$6.115 \\
0.6& 4.120&$-$11.866& 0.437&  0.132& $-$0.049& $-$0.002
& $-$0.001& $-$7.229 \\
0.7& 4.120&$-$13.770& 0.379&  0.130& $-$0.057& $-$0.001
&    0.005& $-$9.194 \\
0.8& 4.120&$-$17.423& 0.608&  0.089& $-$0.069&    0.001
&    0.013&$-$12.661 \\
0.9& 4.120&$-$25.269&1.952&$-$0.085& $-$0.085&    0.005
&    0.017&$-$19.379 \\
\hline
\end{tabular*}
\end{table*}
Each of the contributions to $\sigma$ has a sign that can change
because of the interplay between real and virtual corrections.
The cross--section corresponding to  the Born diagrams for producing
a real particle is always positive, whereas
the sign of the radiative corrections depends on the order of
perturbation theory. For the virtual corrections at odd orders it is
negative, and at even orders it is positive.
When the aperture of the counters is small the compensation between real
and virtual corrections is not complete.
In the limiting case of small aperture $(\rho\to 1,\ x_c\to 1)$
the virtual contributions dominate.
\vskip 1.pt
The numerical results were obtained by using the NLLBHA fortran code 
\cite{nllbha}. 
\vskip 1.pt
The approach described above to the small angle electron-positron 
cross-section can be also used to evaluate, with a next-to-next-to-leading
accuracy, radiative corrections to the electron line at HERA in the
small-$x$ region. This problem is under investigation. 
\vskip 1.pt
The analytical and the numerical calculations for the cross--section
in the non symmetrical Narrow-Wide configuration are in progress and
will be presented elsewhere.
\vskip 10.0pt

\leftline{\bf Acknowledgements}
\vskip 10.0pt 
We thank Johannes Bluemlein for inviting us
to this workshop.
We are grateful for support to the Istituto Nazionale 
di Fisica Nucleare (INFN), to the International Association (INTAS) 
for the grant 93-1867 and 
to the Russian Foundation for Basic Research (RFBR)
for the grant 96-02-17512.
One of us (L.T.) would like to thank H.~Czyz, M.~Dallavalle,
B.~Pietrzyk and T.~Pullia for several useful discussions at various stages
of the work and the CERN theory group for the hospitality. 
One of us (A.A.) is thankful to the Royal Swedish Academy of Sciences for
an ICFPM grant.

\end{document}